\documentstyle[abe,ijmp,12pt]{article}
\begin{document}
\input mssymb.tex
\pagestyle{empty}
\baselineskip20pt
\font\hugeit=cmti10 scaled \magstep4
%%\jfont\eightdm=dm10
%%\def\smallat{{\eightdm{}¡÷}}
\def\today{\ifcase\month\or
  January\or February\or March\or April\or May\or June\or
  July\or August\or September\or October\or November\or December\fi
%%  \space\number\day, 
  \ \ \number\year}
\vspace*{-40pt}
\rightline{\bf RIMS-1164}
\vspace*{30pt}
\centerline{\cmssB Question on {\hugeit D\/}=26}
\vskip10pt
\centerline{\cmssB --- String Theory {\hugeit versus\/} Quantum Gravity ---}
\vskip50pt
\centerline{
 Mitsuo Abe\foot(*,{
%%  E-mail: abe{\smallat}kurims.kyoto-u.ac.jp}) 
  E-mail: abe@kurims.kyoto-u.ac.jp}) 
 }
%\vskip3pt
\centerline{\it Research Institute for Mathematical Sciences,
Kyoto University, Kyoto 606-01, Japan}
\vskip3pt
\centerline{ and }
\vskip3pt
\centerline{
 Noboru Nakanishi\foot({\dagger},{
%%  E-mail: nbr-nakanishi{\smallat}msn.com})}
  E-mail: nbr-nakanishi@msn.com})}
%\vskip3pt
\centerline{\it 12-20 Asahigaoka-cho, Hirakata 573, Japan}
\vskip20pt
%% \centerline{ \today }

\vskip100pt
\centerline{\bf Abstract}
In the covariant-gauge two-dimensional quantum gravity, various
derivations of the critical dimension  $D=26$ of the bosonic string
are critically reviewed, and their interrelations are clarified.
It is shown that the string theory is not identical with the proper
framework of the two-dimensional quantum gravity, but the former
should be regarded as a particular aspect of the latter.
The appearance of various anomalies is shown to be explainable 
in terms of a new type of anomaly in a unified way.

\vskip20pt
%\noindent{{\it PACS:\/} 11.25.-w; 04.60.Kz}\hfill\break
%\noindent{{\it Keywords:\/} bosonic string; conformal anomaly; 
%gauge independence}

%%11.25.-w  Theory of fundamental strings
%%03.70.+k  Theory of quantized fields
%%          (see also 11.10 Field theory)
%%04.60.Kz  Lower dimensional models; minisuperspace models
%%11.10.Kk  Field theories in dimensions other than four
%%          (see also 04.50 Gravity in more than four dimensions; 04.60.K
%%          Lower dimensional models in quantum gravity)

\vfill\eject

\pagestyle{plain}
\setlength{\oddsidemargin}{.5truecm}
\setlength{\textheight}{23.cm}  %{8.85in}  
\setlength{\textwidth}{16.cm}
\setlength{\topmargin}{-.5cm}
\setlength{\baselineskip}{19.8pt}
\setlength{\parindent}{25pt}
\textfont0=\tenrm  \textfont1=\teni \textfont2=\tensy \textfont3=\tenex
\def\rm{\fam0 \tenrm} \def\mit{\fam1 } \def\cal{\fam2 }
\def\bf{\tenbf}  \def\it{\tenit} \def\sl{\tensl}
\scriptfont0=\sixrm  \scriptfont1=\sixi  \scriptfont2=\sixsy
\scriptscriptfont0=\smallr \scriptscriptfont1=\smalli 
                           \scriptscriptfont2=\smallsy

\rm
%%%%%%%%%%%%%%%%%%%%%%%%%%%%  Section 1  %%%%%%%%%%%%%%%%%%%%%%%%%%%
\Sec{Introduction}
The bosonic string theory can be described by the two-dimensional
quantum gravity coupled with $D$ scalar fields, where $D$ is the 
dimension of the world in which a string lives.
If one calculates the conformal anomaly in the noncovariant-gauge
(e.g., conformal-gauge) two-dimensional quantum gravity, one finds
that it is proportional to $D-26$.  Hence the conformal anomaly is 
absent if and only if $D=26$.  Thus the two-dimensional quantum
gravity reproduces the critical dimension of the bosonic string
of finite length.
\par
About one decade ago, D\"usedau\cite{Dusedau} and several other
authors$^{2-6}$
%\cite{Baulieu-Bilal,Rebhan-Kraemmer,Kraemmer-Rebhan,Lattore, Freedman-Lattore-Pilch} 
calculated the conformal anomaly in the 
covariant-gauge two-dimensional quantum gravity in the framework 
of perturbation theory.  They all claimed that the conformal anomaly is 
proportional to $D-26$ also in the covariant-gauge cases.  
Especially, Kraemmer and Rebhan\cite{Kraemmer-Rebhan} considered
a large class of gauge fixings and claimed the gauge independence of the 
conformal anomaly.
\par
On the other hand, the present authors\cite{AN1} found that D\"usedau's
result\cite{Dusedau} of the conformal anomaly is {\it not\/} necessarily
obtained if we make field redefinition before applying perturbation 
theory.  That is, the way of calculating the conformal anomaly is
ambiguous and $D=26$ is not the unique result.
\par
Recently, Takahashi\cite{Takahashi} has proposed a new way of deriving
$D=26$ in such a way that it is free of the above ambiguity problem.
He has obtained a BRS anomaly proportional to $D-26$.  He has then 
converted this anomaly into the conformal anomaly by adding the 
conformal degree of freedom to the effective action.
\par
The violation of the BRS invariance contradicts all previous work, 
especially, our exact solution to the de Donder-gauge two-dimensional
quantum gravity, which is completely BRS-invariant.  We have therefore
analyzed why such discrepancy can arise\cite{AN2} and found that 
there is a very delicate problem in evaluating massless Feynman
integrals.  We believe that if there are two regularizations, gauge
invariant and non-invariant, one should adopt the former in the 
calculation of anomaly.  If this principle is accepted, Takahashi's
BRS-violating result must be abandoned, and therefore the subsequent 
derivation of the conformal anomaly is not acceptable.
\par
The problem which we discuss in the present paper is whether or not 
the covariant-gauge two-dimensional quantum gravity can be identified
with the string theory, that is, whether or not the critical dimension
$D=26$ is an indispensable consequence of the two-dimensional quantum
gravity.   Our conclusion is that {\it the string theory is a particular 
aspect of the two-dimensional quantum gravity\/}.  
The proper framework of the latter is free of BRS anomaly, conformal
anomaly, FP-ghost number current anomaly, etc., but one {\it can\/} 
encounter them at one's will.  We clarify why such a paradoxical 
matter happens.
\par
The present paper is organized as follows.  In Sec.\ 2, we critically 
review the various derivations of $D=26$ based on the conformal anomaly
in the covariant-gauge two-dimensional quantum gravity.
In Sec.\ 3, we describe the proper framework of the de Donder-gauge
two-dimensional quantum gravity.  It is quite a healthy theory except
for one peculiar feature, which is called ``field-equation anomaly''.
In Sec.\ 4, it is pointed out that the result reviewed in Sec.\ 2 is 
a consequence of a particular approach to the two-dimensional quantum
gravity.   In Sec.\ 5, Takahashi's calculation of the BRS anomaly is 
reinterpreted in the BRS-invariant framework, and it is shown that the 
reinterpreted one is essentially equivalent to one of the 
Kraemmer-Rebhan class discussed in Sec.\ 2.   Furthermore, we show 
that various anomalies encountered so far are what one can construct
by using the field-equation anomaly.  The final section is devoted to 
discussion.
\vskip50pt
%
%%%%%%%%%%%%%%%%%%%%%%%%%  Section 2  %%%%%%%%%%%%%%%%%%%%%%%%%%%
\Sec{Perturbative calculations of conformal anomaly}
We discuss the conformal anomaly of bosonic string theory in terms of 
covariant-gauge two-dimensional quantum gravity.
The string coordinates are represented by $D$ scalar fields $\phi_M$  
$(M=0,\,1,\,\ldots,\,D-1)$.   We introduce the gravitational field 
$g_{\mu\nu}$, the gravitational B-field $\tilde b_\rho$, the gravitational
FP-ghost $c^\sigma$ and antighost $\bar c_\tau$.  The conformal degree of 
freedom is eliminated so as to avoid the introduction of the Weyl B-field, 
FP-ghost and antighost for simplicity.  We write $g\equiv\det g_{\mu\nu}$ 
and $\tg^{\mu\nu}\equiv(-g)^{1/2}g^{\mu\nu}$; $\det \tg^{\mu\nu}
=-1$ and $\tilde g^{\mu\nu}$ has no conformal degree of freedom.
\par
The conventional BRS transformation is denoted by $\gdel$.  We have
\begin{eqnarray}
&&\gdel g_{\mu\nu}=-\partial_\mu c^\lambda\cdot g_{\lambda\nu}
                   -\partial_\nu c^\lambda\cdot g_{\mu\lambda}
                   -c^\lambda\partial_\lambda g_{\mu\nu}, \\
&&\gdel \tg^{\mu\nu}=\partial_\lambda c^\mu\cdot\tg^{\lambda\nu}
                     +\partial_\lambda c^\nu\cdot\tg^{\mu\lambda}
                     -\partial_\lambda(c^\lambda\tg^{\mu\nu}), \\
&&\gdel c^\sigma=-c^\lambda\partial_\lambda c^\sigma, \\
&&\gdel \bar c_\tau=i\tb_\tau,\\
&&\gdel \tb_\rho=0,\\
&&\gdel \phi_M=-c^\lambda\partial_\lambda\phi_M.
\end{eqnarray}
\par
The Lagrangian density $\lag$ consists of the string one
\begin{eqnarray}
&&\lagS={1\over2}\tg^{\mu\nu}\partial_\mu\phi^M\cdot\partial_\nu\phi_M,
\end{eqnarray}
the gauge-fixing one $\lagGF$ and the FP-ghost one $\lagFP$.
Here $\lagS$ is BRS-invariant and $\lagGF+\lagFP$ is 
BRS-exact.\foot(a,{``BRS-exact''  means that it can be written as $\gdel{(}\ 
\cdot \ {)}$.  Because of $\gdel^2=0$, BRS invariance follows from BRS
exactness.})
In the de Donder gauge, we have
\begin{eqnarray}
&&\lagGF=-\partial_\mu\tb_\nu\cdot\tg^{\mu\nu},\\
&&\lagFP=-i\partial_\mu\bar c_\nu\cdot
  [ \tg^{\mu\lambda}\partial_\lambda c^\nu
   +\tg^{\lambda\nu}\partial_\lambda c^\mu
   -\partial_\lambda(\tg^{\mu\nu}c^\lambda) ].
\end{eqnarray}
\par\vspace*{5pt}
\Subsec{D\"usedau's approach}
D\"usedau\cite{Dusedau} initiated anomaly calculation in the covariant
gauge.  To apply perturbative approach, one sets
\begin{eqnarray}
&&g_{\mu\nu}=\eta_{\mu\nu}+h_{\mu\nu},
\end{eqnarray}
and therefore
\begin{eqnarray}
&&\tg^{\mu\nu}=\eta^{\mu\nu}-\eta^{\mu\sigma}\eta^{\nu\tau}h_{\sigma\tau}
               +{1\over2}\eta^{\mu\nu}\eta^{\sigma\tau}h_{\sigma\tau}
               +\ \cdots.
\end{eqnarray}
Substituting \eqno(2,11) into a sum of \eqno(2,7)--\eqno(2,9), and then 
neglecting higher order terms with respect to quantum fields and also a 
linear total-divergence term, we obtain
\begin{eqnarray}
\lag^{(0)\,\hbox{\sc D}}
&=&{1\over2}\eta^{\mu\nu}\partial_\mu\phi^M\cdot\partial_\nu\phi_M 
  \nonumber\\
&&+\partial_\mu\tb_\nu\cdot(\eta^{\mu\sigma}\eta^{\nu\tau}h_{\sigma\tau}
    -{1\over2}\eta^{\mu\nu}\eta^{\sigma\tau}h_{\sigma\tau})
  \nonumber\\
&&-i\partial_\mu\bar c_\nu\cdot(\eta^{\mu\lambda}\partial_\lambda c^\nu
    +\eta^{\lambda\nu}\partial_\lambda c^\mu
    -\eta^{\mu\nu}\partial_\lambda c^\lambda).
\end{eqnarray}
\par
We introduce a background (i.e., $c$-number) metric $\hat g_{\mu\nu}$
by replacing $\eta_{\mu\nu}$ by $\hat g_{\mu\nu}$ and $\partial_\mu$ by
background-covariant differentiation $\hat\nabla_\mu$ and then by
multiplying the resultant by $(-\hat g)^{1/2}$.   Then \eqno(2,12) becomes
\begin{eqnarray}
\lag^{\hbox{\sc D}}
&=&{1\over2}\tilde{\hat g}{}^{\mu\nu}\partial_\mu\phi^M\cdot\partial_\nu\phi_M
  \nonumber\\
&&+\hat\nabla_\mu\tb_\nu\cdot
   \tilde{\hat g}{}^{\mu\sigma}\hat g^{\nu\tau}
    (h_{\sigma\tau}-{1\over2}\hat g_{\sigma\tau}
    \hat g^{\alpha\beta}h_{\alpha\beta}) \nonumber\\
&&-i\hat\nabla_\mu\bar c_\nu\cdot
  [ \tilde{\hat g}{}^{\mu\lambda}\hat\nabla_\lambda c^\nu
   +\tilde{\hat g}{}^{\lambda\nu}\hat\nabla_\lambda c^\mu
   -\hat\nabla_\lambda(\tilde{\hat g}{}^{\mu\nu}c^\lambda)].
\end{eqnarray}
This expression is a scalar density under general coordinate 
transformations.
As is well known, the Belinfante symmetric energy-momentum tensor is 
given by 
\begin{eqnarray}
&& T^{\hbox{\sc D}}{}_{\mu\nu}=
2(-\hat g)^{-1/2}{\delta\over\delta\hat g^{\mu\nu}}\int d^2x\,
 \lag^{\hbox{\sc D}}\bigg|_{\hat g_{\sigma\tau}=\eta_{\sigma\tau}}.
\end{eqnarray}
Explicitly, we have
\begin{eqnarray}
&&  T^{\hbox{\sc D}}{}_{\mu\nu}=T_{\hbox{\sc S}\, \mu\nu}
  + T^{\hbox{\sc D}}_{\hbox{\sc GF}\, \mu\nu}
  + T^{\hbox{\sc D}}_{\hbox{\sc FP}\, \mu\nu} \\
\noalign{\noindent with}
&&  T_{\hbox{\sc S}\, \mu\nu}=\partial_\mu\phi^M\cdot\partial_\nu\phi_M
                               + \eta_{\mu\nu}\hbox{-term},\\
&&  T^{\hbox{\sc D}}_{\hbox{\sc GF}\, \mu\nu}
     =\partial_\mu\tb_\sigma\cdot h_\nu{}^\sigma
      +\partial_\nu\tb_\sigma\cdot h_\mu{}^\sigma
      +\tb_\sigma\partial^\sigma h_{\mu\nu} + 
      \eta_{\mu\nu}\hbox{-terms}, \\
&&  iT^{\hbox{\sc D}}_{\hbox{\sc FP}\, \mu\nu}
     =[\partial_\mu\bar c_\sigma\cdot\partial_\nu c^\sigma
       +\partial_\sigma(\partial_\mu\bar c_\nu\cdot c^\sigma)
       +\partial_\sigma(\bar c^\sigma\partial_\mu c_\nu) \nonumber \\
&&   \hspace*{50pt}
       -\partial^\sigma\bar c_\mu\cdot\partial_\sigma c_\nu
       + (\mu\leftrightarrow\nu)] + \eta_{\mu\nu}\hbox{-terms}.
\end{eqnarray}
We calculate the two-point function of $T^{\hbox{\sc D}}{}_{\mu\nu}$ 
in the one-loop order.  The free propagators are
\begin{eqnarray}
&&\wightman{\phi^M(x)\phi^N(y)}=\eta^{MN}\Df(x-y),\\
&&\wightman{h_{\mu\nu}(x)\tb_\rho(y)}=
  (\eta_{\mu\rho}\partial_\nu+\eta_{\rho\nu}\partial_\mu
   -\eta_{\mu\nu}\partial_\rho)\Df(x-y),\\
&&\wightman{c^\sigma(x)\bar c_\tau(y)}=i\delta^\sigma{}_\tau\Df(x-y),
\end{eqnarray}
where $\wightman{\,\cdots\,}$ and $\Df$ denote the vacuum expectation 
value of a time-ordered product and a massless Feynman propagator, 
respectively.  The one-loop formulae
\begin{eqnarray}
&&\partial_\mu\partial_\nu\Df(x-y)\cdot\partial_\lambda\partial_\rho\Df(x-y)
  ={1\over2}\vPhi_{\mu\nu\lambda\rho}(x-y) + \cdots,\\
&&\partial_\mu\partial_\nu\partial_\lambda\Df(x-y)\cdot\partial_\rho\Df(x-y)
  =\vPhi_{\mu\nu\lambda\rho}(x-y) + \cdots,
\end{eqnarray}
are derived by using dimensional regularization, where
\begin{eqnarray}
&&\vPhi_{\mu\nu\lambda\rho}(\xi)\equiv{i\over12\pi}\int 
{d^2p\over(2\pi)^2}\, {p_\mu p_\nu p_\lambda p_\rho \over p^2+i0}
e^{-ip\xi}
\end{eqnarray}
and dots indicate (divergent) local terms.  We then obtain
\begin{eqnarray}
&&\wightman{T^{\hbox{\sc D}}{}_{\mu\nu}(x)
            T^{\hbox{\sc D}}{}_{\lambda\rho}(y)}
   =(D+26-52)\vPhi_{\mu\nu\lambda\rho}(x-y) + \cdots.
\end{eqnarray}
The conformal-anomaly term proportional to $\vPhi_{\mu\nu\lambda\rho}$
vanishes if and only if $D=26$.   Thus the critical dimension is found
to be $D=26$.
\eject  %\par\vspace*{5pt}
\Subsec{Baulieu-Bilal's approach}
Since there is an interesting identity
\begin{eqnarray}
&&\tilde{\hat g}{}^{\mu\lambda}\hat\nabla_\lambda c^\nu
  +\tilde{\hat g}{}^{\lambda\nu}\hat\nabla_\lambda c^\mu
  -\hat\nabla_\lambda(\tilde{\hat g}{}^{\mu\nu}c^\lambda)\nonumber\\
&&\ \ = \tilde{\hat g}{}^{\mu\lambda}\partial_\lambda c^\nu
       +\tilde{\hat g}{}^{\lambda\nu}\partial_\lambda c^\mu
       -\partial_\lambda(\tilde{\hat g}{}^{\mu\nu}c^\lambda),
\end{eqnarray}
the FP-ghost term of $\lag^{\hbox{\sc D}}$ in \eqno(2,13) is quite 
similar to the exact expression for $\lagFP$ in \eqno(2,9) if 
$\hat g_{\mu\nu}$ is replaced by $g_{\mu\nu}$.  
This similarity is complete in the case of conformal gauge because
$\partial_\mu\bar c_\nu$ is replaced by $\bar c_{\mu\nu}$.  
This fact has induced some confusion.
\par
Baulieu and Bilal\cite{Baulieu-Bilal} introduced a background metric,
but it was inessential; their Lagrangian density is essentially 
nothing but the exact one:
\begin{eqnarray}
&&\lag^{\hbox{\sc BB}}{}_{\hbox{\sc GF}}
 =\tb_\mu\partial_\nu\tg^{\mu\nu}, \\
&&\lag^{\hbox{\sc BB}}{}_{\hbox{\sc FP}}
 =-i\partial_\mu\bar c_\nu\cdot(\tg^{\mu\lambda}\partial_\lambda c^\nu
   -\partial_\lambda\tg^{\nu\lambda}\cdot c^\mu),
\end{eqnarray}
where $\lag^{\hbox{\sc BB}}{}_{\hbox{\sc GF}}-\lagGF$  and 
$\lag^{\hbox{\sc BB}}{}_{\hbox{\sc FP}}-\lagFP$ are total divergences.
They eliminated $\lag^{\hbox{\sc BB}}{}_{\hbox{\sc GF}}$ by substituting
the field equation $\partial_\nu\tg^{\mu\nu}=0$  (This is an 
unjustifiable procedure).  Then they considered the second 
variational-derivative of the effective action $\Gamma$ with respect
to the {\it quantum\/} gravitational field, that is, the two-point
function which they considered is 
\begin{eqnarray}
&&\wightman{\calT^{\hbox{\sc BB}}{}_{\mu\nu}(x)
            \calT^{\hbox{\sc BB}}{}_{\lambda\rho}(y)}
\end{eqnarray}
where we use, in general, a script letter ``$\,\calT\,$'' for the 
derivative with respect to the {\it quantum\/} field $g_{\mu\nu}$,
that is,
\begin{eqnarray}
&&\calT_{\mu\nu}=2(-g)^{-1/2}{\delta\over\delta g^{\mu\nu}}
   \int d^2x\, \lag
\end{eqnarray}
generally.   It should satisfy
\begin{eqnarray}
&&\calT_{\mu\nu}=0
\end{eqnarray}
as an {\it exact\/} field equation%
% in contrast with the case of conformal gauge
.
\par
Now, the lowest-order expression for $\calT^{\hbox{\sc BB}}{}_{\mu\nu}$ is
given by
\begin{eqnarray}
\calT^{\hbox{\sc BB}\,(0)}{}_{\mu\nu}
&=&\partial_\mu\phi^M\cdot\partial_\nu\phi_M
  -i[\partial_\mu\bar c_\sigma\cdot\partial_\nu c^\sigma
     +\partial_\mu(\partial_\sigma\bar c_\nu\cdot c^\sigma) 
     + (\mu\leftrightarrow\nu)] \nonumber \\
&& + \eta_{\mu\nu}\hbox{-terms}.
\end{eqnarray}
From \eqno(2,32), the coefficient of $\vPhi_{\mu\nu\lambda\rho}$ 
becomes $D-28$,\cite{Kraemmer-Rebhan,AN1} though they claimed $D-26$.
\eject % \par\vspace*{5pt}
\Subsec{Kraemmer-Rebhan's approach}
Kraemmer and Rebhan\cite{Kraemmer-Rebhan} discussed a class of gauge
fixings by extending Rebhan-Kraemmer's example.\cite{Rebhan-Kraemmer}
Although they explicitly considered conformal degree of freedom, 
we here employ the form in which conformal degree of freedom 
is already eliminated.
\par
Their approach starts with
\begin{eqnarray}
&& g_{\mu\nu}=\hat g_{\mu\nu}+h_{\mu\nu}
\end{eqnarray}
instead of \eqno(2,10), that is, the background metric is directly 
introduced.  Accordingly, background general covariance is {\it not\/}
required.  The linearized BRS transform of $h_{\mu\nu}$ is obtained by 
substituting \eqno(2,33) into \eqno(2,1) and by neglecting higher-order
terms with respect to quantum fields; we have
\begin{eqnarray}
&&\gdel h_{\mu\nu}=-\hat g_{\lambda\nu}\partial_\mu c^\lambda
                   -\hat g_{\mu\lambda}\partial_\nu c^\lambda
                   -\partial_\lambda\hat g_{\mu\nu}\cdot c^\lambda.
\end{eqnarray}
\par
Since only one-loop order is relevant, the gauge-fixing term is chosen
to be linear in $h_{\mu\nu}$ from the outset, and moreover the 
zeroth-order term is also neglected because it does not contribute in
lowest order.  Then the gauge-fixing term can generally be written as
\begin{eqnarray}
&&\lag^{\hbox{\sc KR}}{}_{\hbox{\sc GF}}=
 (F_1{}^{\mu\nu\sigma\tau}\partial_\mu\tb_\nu 
 + F_2{}^{\nu\sigma\tau}\tb_\nu)h_{\sigma\tau},
\end{eqnarray}
where $F_1{}^{\mu\nu\sigma\tau}$ and $F_2{}^{\nu\sigma\tau}$ are 
arbitrary functions of $\hat g_{\lambda\rho}$ and 
$\partial_\kappa\hat g_{\lambda\rho}$, but they must satisfy
\begin{eqnarray}
&&F_1{}^{\mu\nu\sigma\tau}\hat g_{\sigma\tau}=0, \qquad
  F_2{}^{\nu\sigma\tau}\hat g_{\sigma\tau}=0
\end{eqnarray}
because conformal degree of freedom is  eliminated.  
The linearized BRS invariance of 
$\lag^{\hbox{\sc KR}}{}_{\hbox{\sc GF}}+\lag^{\hbox{\sc 
KR}}{}_{\hbox{\sc FP}}$ yields
\begin{eqnarray}
&&\lag^{\hbox{\sc KR}}{}_{\hbox{\sc FP}}
=-i(F_1{}^{\mu\nu\sigma\tau}\partial_\mu\bar c_\nu + 
F_2{}^{\nu\sigma\tau}\bar c_\nu)(\hat\nabla_\sigma c_\tau
+\hat\nabla_\tau c_\sigma),
\end{eqnarray}
where use has been made of an identity
\begin{eqnarray}
&&\hat g_{\lambda\nu}\partial_\mu c^\lambda
 +\hat g_{\mu\lambda}\partial_\nu c^\lambda
 +\partial_\lambda\hat g_{\mu\nu}\cdot c^\lambda
 =\hat\nabla_\mu c_\nu+\hat\nabla_\nu c_\mu
\end{eqnarray}
with $c_\mu\equiv\hat g_{\mu\nu}c^\nu$.
\par
Kraemmer and Rebhan defined ``energy-momentum tensor'' $T^{\hbox{\sc 
KR}}{}_{\mu\nu}$ analogously to \eqno(2,14), though, in general,
$T^{\hbox{\sc KR}}{}_{\mu\nu}$ is no longer conserved.  
The coefficient of $\vPhi_{\mu\nu\lambda\rho}$ in 
$\wightman{T^{\hbox{\sc KR}}{}_{\mu\nu}T^{\hbox{\sc 
KR}}{}_{\lambda\rho}}$ is seen to be $D-26$ for various concrete 
examples.  Kraemmer and Rebhan\cite{Kraemmer-Rebhan} claimed that this
is true in general, but their proof cannot be regarded as correct.
According to them, the gauge independence of the anomaly term would be 
a simple consequence of the linearized BRS exactness of the variation
of the effective action under the infinitesimal variation of 
gauge fixing.  But if the same reasoning were applied to the effective
action itself, the anomaly coefficient would be shown to be $D$ 
instead of $D-26$.  The reason why their proof is wrong is the 
{\it noncommutativity\/} between $\gdel$ and 
$\delta/\delta\hat g^{\mu\nu}$, which is seen from \eqno(2,34) and the 
requirement
\begin{eqnarray}
&&{\delta h_{\sigma\tau}\over\delta\hat g^{\mu\nu}}=0.
\end{eqnarray}
As we show in a separate paper\cite{AN3}, however, the gauge 
independence of the anomaly term is correct, as long as 
$\lag^{\hbox{\sc KR}}{}_{\hbox{\sc GF}}$ reduces to the linearized de
Donder gauge fixing (second line of \eqno(2,12)) in the flat limit
$\hat g_{\mu\nu}=\eta_{\mu\nu}$.  The proof is given by directly 
analyzing $\wightman{T^{\hbox{\sc KR}}{}_{\mu\nu}
 T^{\hbox{\sc KR}}{}_{\lambda\rho}}$ in detail.  Kraemmer and Rebhan
claimed their proposition with more generality; for example, according
to them, conformal gauge would be also included.  It is very difficult,
however, to compare generally the two cases having different Feynman
rules in the direct analysis of $\wightman{T^{\hbox{\sc KR}}{}_{\mu\nu}
T^{\hbox{\sc KR}}{}_{\lambda\rho}}$.
\par\vspace*{5pt}
\Subsec{Kraemmer-Rebhan's version of Baulieu-Bilal's approach}
Kraemmer-Rebhan's general treatment\cite{Kraemmer-Rebhan} includes 
D\"usedau's case,\cite{Dusedau} Rebhan-Kraemmer's 
example,\cite{Rebhan-Kraemmer} etc.   Very interesting is the fact that
Baulieu-Bilal's analysis\cite{Baulieu-Bilal} can be reinterpreted as an
example of Kraemmer-Rebhan's framework.\cite{Kraemmer-Rebhan}
\par
The Lagrangian density $\lag^{\hbox{\sc KRBB}}$ is defined by replacing
$\hat\nabla_\mu\tb_\nu$ and $\hat\nabla_\mu\bar c_\nu$ of 
$\lag^{\hbox{\sc D}}$ in \eqno(2,13) by $\partial_\mu\tb_\nu$ and 
$\partial_\mu\bar c_\nu$, respectively.  Owing to the correspondence 
noted in Subsec.\ 2.2, $\lag^{\hbox{\sc KRBB}}$ has the same form as
$\lag^{\hbox{\sc BB}}$, except for the gauge-fixing term, if 
$g_{\mu\nu}$ is replaced by $\hat g_{\mu\nu}$.
\par
Calculating $T^{\hbox{\sc KRBB}}{}_{\mu\nu}$, we obtain
\begin{eqnarray}
T^{\hbox{\sc KRBB}}{}_{\mu\nu}
&=&\partial_\mu\phi^M\cdot\partial_\nu\phi_M +\partial^\sigma\tb_\sigma\cdot 
h_{\mu\nu} \nonumber \\
&&-i[\partial_\mu\bar c_\sigma\cdot\partial_\nu c^\sigma
     +\partial_\mu(\partial_\sigma\bar c_\nu\cdot c^\sigma)
     +(\mu\leftrightarrow\nu)] \nonumber \\
&&+\eta_{\mu\nu}\hbox{-terms}.
\end{eqnarray}
Note that \eqno(2,40) reproduces \eqno(2,32), though it contains the 
contribution from the gauge-fixing term.  The coefficient of the 
conformal-anomaly term is now $D+2-28$ in conformity with the 
Kraemmer-Rebhan's proposition.\cite{Kraemmer-Rebhan}
\vskip50pt
%
%%%%%%%%%%%%%%%%%%%%%%%%%%%%  Section 3   %%%%%%%%%%%%%%%%%%%%%%%%%%
\Sec{Exact treatment}
All discussions made in Sec.\ 2 are based on perturbation theory.
But one should note that it is quite artificial to apply perturbative
approach to the two-dimensional quantum gravity because it contains
no expansion parameter.  One may say that perturbation theory is a loop
expansion rather than a parameter expansion, but one must recognize the
fact that the division of the action into its free part and its interaction
one is artificial and {\it nonunique\/}.  It is discussed in next section
that this fact is really troublemaking in the perturbative approach to 
the conformal anomaly.
\par
Even apart from the string theory, the de Donder gauge two-dimensional
quantum gravity is a very interesting model.  It can be regarded as 
the {\it two-dimensional version of the quantum Einstein gravity\/}
because the two-dimensional Einstein-Hilbert action is trivial.
More importantly, the de Donder-gauge two-dimensional quantum gravity
is, apart from its dimensionality, nothing but the {\it zeroth-order
approximation of the de Donder-gauge quantum Einstein gravity\/}\cite{AN4} 
in the expansion in powers of the Einstein gravitational 
constant $\kappa$ [ The conventional perturbative approach to quantum
Einstein gravity starts with a wrong zeroth order.\cite{N1}].
Furthermore, the two-dimensional quantum gravity is {\it exactly
solvable\/}.
\par
Before entering into the discussion of the two-dimensional quantum gravity,
we briefly review the covariant operator formalism of quantum Einstein
gravity\cite{N2,NO} in the $n$-dimensional spacetime.
\par
First, we introduce the notion of the {\it intrinsic\/} BRS 
transformation,\cite{N3} which is denoted by $\bdel$.  
Just like the angular momentum, the BRS transformation $\gdel$ 
consists of its intrinsic part $\bdel$ and its orbital part:
\begin{eqnarray}
&&\gdel\vPhi=\bdel\vPhi-c^\lambda\partial_\lambda\vPhi,
\end{eqnarray}
where $\vPhi$ is {\it any\/} quantum field.  If $\vPhi$ is a tensor
field, $\bdel\vPhi$ is determined by its general linear transformation
property, e.g.,
\begin{eqnarray}
&&\bdel g_{\mu\nu}=-\partial_\mu c^\lambda\cdot g_{\lambda\nu}
                   -\partial_\nu c^\lambda\cdot g_{\mu\lambda},\\
&&\bdel \phi_M=0.
\end{eqnarray}
On the other hand, for ghost fields, we have
\begin{eqnarray}
&&\bdel c^\sigma=0,\\
&&\bdel \bar c_\tau=ib_\tau,\\
&&\bdel b_\rho=0.
\end{eqnarray}
Just as $\gdel$ is, $\bdel$ is nilpotent, but $\bdel$ does not commute
with $\partial_\mu$: $[\bdel,\;\partial_\mu]=-\partial_\mu 
c^\lambda\partial_\lambda$.  Substituting \eqno(3,5) and \eqno(3,6) into
\eqno(3,1), we have
\begin{eqnarray}
&&\gdel\bar c_\tau=ib_\tau-c^\lambda\partial_\lambda\bar c_\tau,\\
&&\gdel b_\rho=-c^\lambda\partial_\lambda b_\rho.
\end{eqnarray}
The B-field $b_\rho$ is different from the $\tb_\rho$ of Sec.\ 2.
If we set\cite{N4}
\begin{eqnarray}
&&\tb_\rho\equiv b_\rho+ic^\lambda\partial_\lambda\bar c_\rho,
\end{eqnarray}
we find that the conventional rules \eqno(2,1)--\eqno(2,6) of $\gdel$ 
are reproduced.  Thus the effect of considering $\bdel$ is to regard
$b_\rho$, rather than $\tb_\rho$, as the primary B-field.  
Contrary to its first impression, this problem is not so trivial.
\par
Just as in constructing a Lorentz-invariant action, it is conceptually
simper to require the {\it intrinsic\/} BRS invariance of the action,
that is,\cite{N3}
\begin{eqnarray}
&&\bdel[(-g)^{-1/2}\lag]=0.
\end{eqnarray}
The Lagrangian density $\lag$ of the quantum Einstein gravity is 
given by
\begin{eqnarray}
&&\lag=\lag_{\hbox{\sc E}}+\lagGF+\lagFP+\lagm
\end{eqnarray}
with
\begin{eqnarray}
&&\lag_{\hbox{\sc E}}={1\over2\kappa}(-g)^{1/2}R,
\end{eqnarray}
$R$ being the scalar curvature,
\begin{eqnarray}
&&\lagGF=-\tg^{\mu\nu}\partial_\mu b_\nu,\\
&&\lagFP=-i\tg^{\mu\nu}\partial_\mu\bar c_\sigma\cdot\partial_\nu 
c^\sigma,
\end{eqnarray}
and if we consider free scalar fields $\phi_M$ as matter, 
$\lagm=\lagS$.  Compared with \eqno(2,9), the simplicity of \eqno(3,14)
is quite remarkable.  The appearance of the simple derivative for
$c^\sigma$ is the manifestation of the abelian nature of translation
group which is the global version of general coordinate transformation
group.
\par
The field equations which follow from \eqno(3,11) are
\begin{eqnarray}
&&R_{\mu\nu}-{1\over2}g_{\mu\nu}R
 =\kappa(E_{\mu\nu}-{1\over2}g_{\mu\nu}g^{\sigma\tau}E_{\sigma\tau}
 -T_{\hbox{\sc S}\,\mu\nu}) \\
\noalign{\noindent with $R_{\mu\nu}$ being the Ricci tensor and}
&&E_{\mu\nu}\equiv\partial_\mu b_\nu+i\partial_\mu\bar 
c_\sigma\cdot\partial_\nu c^\sigma+(\mu\leftrightarrow\nu),\\
&&\partial_\mu\tg^{\mu\nu}=0,\\
&&\partial_\mu(\tg^{\mu\nu}\partial_\nu c^\sigma)=0,\\
&&\partial_\mu(\tg^{\mu\nu}\partial_\nu \bar c_\tau)=0,\\
&&\partial_\mu(\tg^{\mu\nu}\partial_\nu \phi_M)=0.
\end{eqnarray}
Taking the covariant derivative of \eqno(3,15) and using 
\eqno(3,17)--\eqno(3,19), we obtain
\begin{eqnarray}
&&\partial_\mu(\tg^{\mu\nu}\partial_\nu b_\rho)=0.
\end{eqnarray}
This beautiful result holds only for $b_\rho$ but not for $\tb_\rho$.
\par
Introducing a $(4n+D)$-dimensional ``supercoordinate''\cite{N5}
\begin{eqnarray}
&& X=\{\,x^\lambda,\ b_\rho,\ c^\sigma,\ \bar c_\tau,\ \phi_M\,\},
\end{eqnarray}
we can summarize \eqno(3,17)--\eqno(3,21) as 
\begin{eqnarray}
&&\partial_\mu(\tg^{\mu\nu}\partial_\nu X)=0.
\end{eqnarray}
From \eqno(3,23), we see that the $(4n+D)$-dimensional ``supermomentum''
\begin{eqnarray}
&&{\cal P}^\mu(X)\equiv\tg^{\mu\nu}\partial_\nu X  \\
\noalign{\noindent and ``angular supermomentum''}
&&{\cal M}^\mu(X,\,Y)\equiv
\tg^{\mu\nu}(X\partial_\nu Y-\partial_\nu X\cdot Y)
\end{eqnarray}
are conserved, where $Y$ denotes another supercoordinate.
After canonical quantization, the corresponding charge operators form
$(4n+D)$-dimensional Poincar\'e-like superalgebra 
$IO\!S\!p(2n+D,\,2n)$.\cite{N5}$^{\hbox{\sc ,}}$\cite{N2,NO}  
The existence of this remarkable
symmetry justifies the assertion that $b_\rho$, but not $\tb_\rho$, 
should be regarded as the primary field.
\par
Now, we proceed to considering the two-dimensional case 
$n=2$.\cite{AN5,A}   As is well known, in this case, $\lag_{\hbox{\sc 
E}}$ is a total divergence and should be omitted.  Then $\lag$ 
reduces to
\begin{eqnarray}
&&\lag=-{1\over2}\tg^{\mu\nu}(E_{\mu\nu}
     -\partial_\mu\phi^M\cdot\partial_\nu\phi_M).
\end{eqnarray}
Because of $\det \tg^{\mu\nu}=-1$, \eqno(3,26) contains only two 
degrees of freedom of $g_{\mu\nu}$.  We can directly deal with 
\eqno(3,26), but it is not convenient to do so.  More convenient
is to add the conformal degree of freedom to \eqno(3,26) and then extract 
the part independent of the conformal degree of freedom; the results are 
the same.
So, we add $(-g)^{1/2}R\,b$ to \eqno(3,26).  After deriving field
equations and equal-time commutation relations, we set $b=0$.
\par
The field equations are\foot(b,{$\tg_{\mu\nu}$ is the inverse of
$\tg^{\mu\nu}$.})
\begin{eqnarray}
&&\calT_{\mu\nu}\equiv T_{\hbox{\sc S}\,\mu\nu}-E_{\mu\nu}
  +{1\over2}\tg_{\mu\nu}\tg^{\sigma\tau}E_{\sigma\tau}=0
\end{eqnarray}  
and \eqno(3,17)--\eqno(3,20).  Since \eqno(3,21) is again derived,
we have the $IO\!S\!p(4+D,\,4)$ symmetry.   It should be noted that 
\eqno(3,27) is nothing but \eqno(2,31).
\par
It is quite remarkable that all two-dimensional commutation relations 
are explicitly obtained in closed form if we introduce the $q$-number
D function, $\calD(x,\,y)$, defined by a $q$-number Cauchy problem,
which involves $\tg^{\mu\nu}$ only.  Some of the exact two-dimensional
commutation relations are as follows:
\begin{eqnarray}
&&\{c^\sigma(x),\;\bar c_\tau(y)\}=-\delta^\sigma{}_\tau\calD(x,\,y),\\
&&[\tg^{\mu\nu}(x),\;b_\rho(y)]=i\{\gdel[\tg^{\mu\nu}(x)],\;\bar 
c_\rho(y)\},\\
&&[b_\lambda(x),\;b_\rho(y)]=i[\partial_\rho 
b_\lambda(x)+\partial_\lambda b_\rho(y)]\cdot\calD(x,\,y)
\end{eqnarray}
and $\tg^{\mu\nu}(x)$ commutes with any field but $b_\rho(y)$.
Note that all those relations are form-independent of matter field.
From the two-dimensional commutation relations, we can explicitly 
calculate all multiple commutators in closed form.
\par
The representation of the above field algebra is constructed by giving
all Wightman functions (i.e., vacuum expectation values of simple 
product of fields).$^{20-24}$ %\cite{AN6,AN7,AN8,AN9,AN10}
With one-point Wightman functions
\begin{eqnarray}
&&\wightman{\tg^{\mu\nu}(x_1)}_{\hbox{\sc W}}=\eta^{\mu\nu}, \qquad
\wightman{\vPhi(x_1)}_{\hbox{\sc W}}=0,
\end{eqnarray}
where $\vPhi$ denotes any field other than $\tg^{\mu\nu}$, $N$-point
Wightman functions $W(x_1,\,\cdots\,,\,x_N)$ are constructed so as to 
satisfy the following requirements.
\begin{item1}
\item[\bf W1] They are consistent with the vacuum expectation values 
of all multiple commutators.
\item[\bf W2] [Energy-positivity requirement] They are boundary 
values of analytic functions of $x_1{}^\mu-x_2{}^\mu,\,\cdots\,,\,
x_{N-1}{}^\mu-x_N{}^\mu$ from the lower half-planes of their $\mu=0$
components.
\item[\bf W3] [Generalized normal-product rule] $W(x_1,\cdots,x_N)$
with $x_i{}^\mu=x_{i+1}{}^\mu=\cdots=x_j{}^\mu$ $(i<j)$ is defined 
from $W(x_1,\cdots,x_N)$ by setting 
$x_i{}^\mu=x_{i+1}{}^\mu=\cdots=x_j{}^\mu$ and by deleting the 
resulting divergent terms in such a way that it be independent of 
the ordering of $i,\,i+1,\,\ldots,\,j$.
\end{item1}
\par
Because of the self-commutativity of $\tg^{\mu\nu}$, we have
\begin{eqnarray}
&&\wightman{F(\tg^{\mu\nu})}_{\hbox{\sc W}}=F(\eta^{\mu\nu}),
\end{eqnarray}
for any function $F$ of $\tg^{\mu\nu}$.  Especially, we have
\begin{eqnarray}
&&\wightman{\calD(x,\,y)}=D(x-y),
\end{eqnarray}
which is the usual two-dimensional D-function.  It is decomposed into
positive-energy and negative-energy parts:
\begin{eqnarray}
&&iD(x-y)=\Dp(x-y)-\Dp(y-x).
\end{eqnarray}
According to {\bf W2}, $W(x_1,\cdots,x_N)$ is expressed only in terms 
of $\Dp(x_i-x_j)$ $(i<j)$.
\par
All Wightman functions are explicitly 
constructed.$^{20-24}$
%\cite{AN6,AN7,AN8,AN9,AN10}  
We here present some examples of them:
\begin{eqnarray}
&&\wightman{b_\lambda(x_1)b_\rho(x_2)}_{\hbox{\sc W}}
 =\partial_\lambda{}^{x_2}\Dp(x_1-x_2)\cdot
  \partial_\rho{}^{x_1}\Dp(x_1-x_2), \\
&&\wightman{\tg^{\mu\nu}(x_1)b_\rho(x_2)}_{\hbox{\sc W}}
 =-(\delta^\mu{}_\rho\partial^\nu+\delta^\nu{}_\rho\partial^\mu
    -\eta^{\mu\nu}\partial_\rho)^{x_1}
   \Dp(x_1-x_2), \\
&&\wightman{c^\sigma(x_1)\bar c_\tau(x_2)b_\rho(x_3)}_{\hbox{\sc W}}
 =i\delta^\sigma{}_\tau[\partial_\rho{}^{x_1}\Dp(x_1-x_2)\cdot\Dp(x_1-x_3)
   \nonumber \\
&&\hspace*{150pt} +\partial_\rho{}^{x_2}\Dp(x_1-x_2)\cdot\Dp(x_2-x_3)].
\end{eqnarray}
The corresponding $\tau$-functions (i.e., vacuum expectation values 
of time-ordered products) are obtained by simply replacing $\Dp$ by
$\Df$.
\par
Our exact solution is consistent with the field equations 
\eqno(3,17)--\eqno(3,21) and with all symmetry generators of 
$IO\!S\!p(4+D,4)$, provided that they are constructed from \eqno(3,24)   
and \eqno(3,25).  However, our exact solution is {\it not\/} 
consistent with the field equation \eqno(3,27).
Indeed,from \eqno(3,35)--\eqno(3,37) and $\wightman{\phi_M 
b_\rho}_{\hbox{\sc W}}=0$, we find that\cite{AN7,AN8}
\begin{eqnarray}
&&\wightman{\calT_{\mu\nu}(x_1)b_\rho(x_2)}_{\hbox{\sc W}}\not=0.
\end{eqnarray}
We call this phenomenon ``field-equation anomaly''.
We note that the violation is very little in the sense that the degree
of freedom of \eqno(3,27) is the same as that of \eqno(3,21).
The existence of field-equation anomaly is demonstrated also in some
simpler models.\cite{AN11,AN12}
\par
Our exact solution is completely BRS invariant.  The corresponding BRS
charge is given by 
\begin{eqnarray}
&&Q_b\equiv \int dx^1\,\tg^{0\nu}(b_\rho\partial_\nu c^\rho
  -\partial_\nu b_\rho\cdot c^\rho).
\end{eqnarray}
However, the BRS {\it Noether\/} charge is anomalous because in order
to reduce it to \eqno(3,39), one has to use $\calT_{\mu\nu}=0$.
\vskip50pt
%
%%%%%%%%%%%%%%%%%%%%%%%%%%%  Section 4  %%%%%%%%%%%%%%%%%%%%%%%%%%%%%%
\Sec{Nonuniqueness of the conformal anomaly}
Now, we go back to the problem of the conformal anomaly.
As discussed in Sec.\ 3, it is more natural to regard $b_\rho$, rather
than $\tb_\rho$, as the primary field in the framework of the 
$n$-dimensional quantum Einstein gravity and therefore in its $n=2$
case.
\par
We first write the gauge-fixing plus FP-ghost part of \eqno(3,26):
\begin{eqnarray}
&&\lagGF+\lagFP=-\tg^{\mu\nu}\partial_\mu b_\nu
 -i\tg^{\mu\nu}\partial_\mu\bar c_\sigma\cdot\partial_\nu c^\sigma.
\end{eqnarray}
We rewrite it by using \eqno(3,9) into
\begin{eqnarray}
\lagGF+\lagFP&=&-\tg^{\mu\nu}\partial_\mu\tb_\nu
-i\tg^{\mu\nu}\partial_\mu(\partial_\lambda\bar c_\nu\cdot c^\lambda) 
-i\tg^{\mu\nu}\partial_\mu\bar c_\sigma\cdot\partial_\nu c^\sigma.
\end{eqnarray}
Comparing \eqno(4,2) with \eqno(2,8) plus \eqno(2,9), we find that 
the difference between them is a total divergence
\begin{eqnarray}
&&i\partial_\lambda(\tg^{\mu\nu}\partial_\mu\bar c_\nu\cdot c^\lambda).
\end{eqnarray}
Both are, therefore, equivalent in the usual sense.  They are {\it not\/},
however, in the problem of the conformal anomaly.\cite{AN1}
\par
The free Lagrangian density which follows from \eqno(3,26) is
\begin{eqnarray}
\lag^{(0)}&=&{1\over2}\eta^{\mu\nu}\partial_\mu\phi^M\cdot\partial_\nu\phi_M
 \nonumber\\
 &&+\partial_\mu b_\nu\cdot(\eta^{\mu\sigma}\eta^{\nu\tau}h_{\sigma\tau}
  -{1\over2}\eta^{\mu\nu}\eta^{\sigma\tau}h_{\sigma\tau}) \nonumber\\
 &&-i\eta^{\mu\nu}\partial_\mu\bar c_\sigma\cdot\partial_\nu c^\sigma,
\end{eqnarray}
which should be compared with \eqno(2,12).  The Feynman propagators 
implied by \eqno(4,4) are the same as those which follows from \eqno(2,12) 
if $\tb_\rho$ is replaced by $b_\rho$.  More generally, we may 
consider\foot(c,{Note that {\eqno(4,5)} contains no second-order 
derivatives.})
\begin{eqnarray}
&&\lag^\alpha=\lag^{(0)}
-i\alpha\partial_\lambda(\eta^{\lambda\nu}\partial_\mu\bar c_\nu\cdot c^\mu
  -\eta^{\mu\nu}\partial_\mu\bar c_\nu\cdot c^\lambda),
\end{eqnarray}
$\alpha$ being an arbitrary constant.  Of course, 
$\lag^{\alpha=0}=\lag^{(0)}$ and $\lag^{\alpha=1}$ is identifiable with
$\lag^{(0)\,\hbox{\sc D}}$.  In spite of the $\alpha$ independence of the
action, the corresponding symmetric energy-momentum tensor 
$T^\alpha{}_{\mu\nu}$ is nontrivially $\alpha$-dependent.  Indeed,
\begin{eqnarray}
iT^\alpha{}_{\hbox{\sc FP}\, \mu\nu}
&=&\partial_\mu\bar c_\sigma\cdot\partial_\nu c^\sigma  
  +{1\over2}\partial_\sigma[-\partial_\mu\bar c^\sigma\cdot c_\nu
  +\bar c^\sigma\partial_\mu c_\nu
  -\bar c_\nu\partial_\mu c^\sigma +\partial_\mu\bar c_\nu\cdot c^\sigma
   + (\mu\leftrightarrow\nu)] \nonumber\\
&&+{1\over2}\alpha\partial_\sigma[\partial_\mu(\bar c^\sigma c_\nu+
  \bar c_\nu c^\sigma)-\partial^\sigma(\bar c_\mu c_\nu) 
  +(\mu\leftrightarrow\nu)] \nonumber\\
&&+\eta_{\mu\nu}\hbox{-terms}.
\end{eqnarray}
We calculate the coefficient of $\vPhi_{\mu\nu\lambda\rho}$ as done in 
Sec.\ 2.   Unfortunately, we generally encounter infrared divergence $I$
arising from $\partial_\mu\partial_\nu\partial_\lambda\partial_\rho\Df\cdot
\Df$.\foot(d,{This fact was overlooked in Ref.\ 7.  Eq.\ {(27)} of that paper
is reproduced if $I$ is replaced by $1/2$.})
We find that
\begin{eqnarray}
&&\wightman{T^\alpha{}_{\mu\nu}T^\alpha{}_{\lambda\rho}}
=[D+26-6-24\alpha-22\alpha^2+4(1-\alpha^2)I]\vPhi_{\mu\nu\lambda\rho}
 + \cdots.
\end{eqnarray}
The conformal anomaly is finite if $\alpha=1$ or $\alpha=-1$.  
Anyway, the important fact is the {\it nonuniqueness\/} of the conformal
anomaly.  Though the nonuniqueness of the symmetric energy-momentum
tensor is contrary to the common sense, one should note that 
$T^\alpha{}_{\mu\nu}$ is {\it not\/} an observable for any value of 
$\alpha$.
\par
Now, we discuss the compatibility between the above observation of the 
nonuniqueness and Kraemmer-Rebhan's gauge independence\cite{Kraemmer-Rebhan}
of the conformal anomaly.  As noted in Subsec.\ 2.2, Kraemmer and Rebhan
employed perturbative approach from the outset, regarding $\tb_\rho$ as 
the primary field as a matter of course.  Within such a perturbative 
framework, the conformal anomaly is unique and the critical dimension is
$D=26$.  However, such a nonlinear transformation of Heisenberg fields as
\eqno(3,9) is beyond their scope.  {\it Field redefinition generally
changes the division of the total Lagrangian density into the free part and
the interaction one\/}.  That is, given a theory, its perturbative 
expansion is {\it not\/} uniquely determined.  Thus the conformal anomaly is 
dependent on the choice of perturbation theory, as long as we define the
conformal anomaly by using such a quantity as $T_{\mu\nu}$.
\vskip50pt
%
%%%%%%%%%%%%%%%%%%%%%%%%%%%%%  Section 5  %%%%%%%%%%%%%%%%%%%%%%%%%%%%
\Sec{Takahashi's anomaly calculation and field-equation anomaly}
Recently, Takahashi\cite{Takahashi} has proposed a new approach to the 
problem of the critical dimension $D=26$.  In order to avoid the ambiguity
problem stated in Sec.\ 4, $T_{\mu\nu}$ is not considered in his approach.
Since his approach is very interesting, we first review it briefly.
\par
He positively admits the transformation \eqno(3,9), i.e.,
\begin{eqnarray}
&&\tb_\rho=-i\gdel(\bar c_\rho)=b_\rho+ic^\sigma\partial_\sigma\bar c_\rho.
\end{eqnarray}
In the theory defined by \eqno(3,26), he calculates
\begin{eqnarray}
\wightman{\tb_\lambda(x)\tb_\rho(y)}
&=&\wightman{b_\lambda(x)b_\rho(y)}
  +i\wightman{b_\lambda(x)c^\sigma(y)\partial_\sigma\bar c_\rho(y)} 
  \nonumber\\
&&+i\wightman{c^\sigma(x)\partial_\sigma\bar c_\lambda(x)\cdot b_\rho(y)}
  \nonumber \\
&&-\wightman{c^\sigma(x)\partial_\sigma\bar c_\lambda(x)\cdot
   c^\tau(y)\partial_\tau\bar c_\rho(y)}
\end{eqnarray}
perturbatively in one-loop approximation, which is actually exact.
It should be noted that the perturbative orders of the first, second,
third, and fourth terms are 2, 1, 1, and 0, respectively.
By employing the lightcone-coordinate method (or dimensional 
regularization), he obtains 
\begin{eqnarray}
&& {\cal F}\wightman{\tb_\lambda(x)\tb_\rho(y)}
 ={D-26\over 48\pi}\,{ip_\lambda p_\rho\over p^2+i0} + \cdots,
\end{eqnarray}
where ${\cal F}$ denotes the Fourier transformation and dots represent 
divergent local terms proportional to $\eta_{\lambda\rho}$.
Because $\tb_\lambda\tb_\rho$ is BRS-exact, 
$\wightman{\tb_\lambda\tb_\rho}$ should vanish unless the BRS invariance 
is anomalous or spontaneously broken.
\par
Takahashi\cite{Takahashi} calculates the effective action by taking the 
inverse of the matrix formed by the two-point functions $\{{\cal 
F}\wightman{h_{\mu\nu}h_{\sigma\tau}}=0,\,{\cal 
F}\wightman{h_{\mu\nu}\tb_\rho},\,{\cal F}\wightman{\tb_\lambda 
h_{\sigma\tau}},\,{\cal F}\wightman{\tb_\lambda\tb_\rho}\}$.
It also contains the nonlocal terms proportional to $D-26$.  He then 
recovers the BRS invariance by explicitly introducing the conformal
degree of freedom.  The nonlocal term is converted into the conformal anomaly.
Indeed, he obtains the action of Polyakov's ``induced'' quantum 
gravity\cite{Polyakov} at the one-loop level, though the theory is no
longer exact at the one-loop level.  Thus he claims that $D=26$ is 
obtained without ambiguity.
\par
It is very surprising that Takahashi has found the violation of the
BRS invariance in the original framework of the two-dimensional
quantum gravity.  So far, nobody has claimed the violation of its
BRS invariance.  Kraemmer-Rebhan's gauge independence of the conformal
anomaly was based on the BRS invariance.  More importantly, as explained
in Sec.\ 3, the exact solution to the de Donder-gauge two-dimensional
quantum gravity is completely consistent with the BRS invariance.  
Takahashi's result is evidently inconsistent with our exact solution.
\par
We have therefore investigated the reason for the discrepancy.\cite{AN2}
In order to concentrate our attention only to resolving this problem,
we compare both results by modifying them into the ones on a background as
common as possible, rather than respecting the original standpoints of his
work and ours.   Concretely, we have made the following.
\begin{item1}
\item[1.]  We extract Takahashi's calculation of \eqno(5,3) only,
neglecting his consideration on the effective action.
\item[2.]  We reproduce our BRS-invariant result by means of perturbation
theory.
\item[3.]  All loop integrals are evaluated by using dimensional 
regularization.
\end{item1}
\par
We have found that the qualitative reason for the discrepancy is the 
ambiguity of the massless Feynman integrals themselves:  Because of the 
presence of both ultraviolet and infrared divergences, there is no analytic
domain of the complex dimension $n$.  The quantitative reason for the 
discrepancy is explained in the following way.
\par
The first term of \eqno(5,2), $\wightman{b_\lambda b_\rho}$, is calculated
in the second-order perturbation theory.  The relevant Feynman diagram 
consists of two internal lines and two external lines.  The loop, of 
course, consists of the former only.  Likewise, $\wightman{b_\lambda\cdot
c^\sigma\partial_\sigma\bar c_\rho}$ and $\wightman{c^\sigma\partial_\sigma
\bar c_\lambda\cdot b_\rho}$ have one external line, in addition to a loop
consisting of two internal lines. If one calculates the loop integrals by 
applying the dimensional method, {\it  keeping the external lines strictly
two-dimensional\/}, then one obtains Takahashi's result \eqno(5,3).
On the other hand, if one does by applying the dimensional method to
{\it all lines including external lines\/}, then one obtains the 
BRS-invariant result.  Thus the discrepancy arises from whether the 
dimension of external lines is 2 or $n\,(\rightarrow2)$.
\par
The reason why the dimension of external lines is relevant is as follows.
The propagator of an external line is a Fourier transform of \eqno(2,20),
that is,
\begin{eqnarray}
&&{\eta_{\mu\rho}p_\nu+\eta_{\nu\rho}p_\mu-\eta_{\mu\nu}p_\rho\over
   p^2+i0}.
\end{eqnarray}
From the loop-integral calculation, we obtain a term proportional to 
$\eta^{\mu\nu}$.  Then the last term of \eqno(5,4) yields a trace of 
$\eta_{\mu\nu}$, which equals the dimension of the external line, i.e.,
2 or $n$.  Usually, since $\vep\equiv n-2\,\rightarrow 0$, the 
difference is trivial.  In the present case, however, the loop integral
consists of a convergent nonlocal term and divergent local terms.
The term proportional to $\eta^{\mu\nu}$, which is one of the latter, is
of order $\vep^{-1}$.  Hence a divergent local term becomes a finite 
nonlocal term if it is multiplied by $\vep p_\rho/(p^2+i0)$.
This explains the discrepancy in the evaluation of $\wightman{\tb_\lambda
\tb_\rho}$.
\par
Takahashi asserts that the right way is to keep external lines strictly 
two-dimensional because the effective action, on which anomaly should be
based, has no external lines.  We cannot agree with him because the 
nonuniform application of the dimensional method violates gauge invariance.
Although Takahashi asserts that his regularization method is consistent 
with the Ward-Takahashi identities for the effective action, their validity
itself is {\it not\/} guaranteed if the Ward-Takahashi identity for the 
$\tau$-functions (Green's functions) are violated. Indeed, it is logically
impossible to obtain a BRS-violating result from a gauge-invariant 
regularization.  We believe that it is natural to apply the dimensional
method {\it to all lines equally so as to keep gauge invariance\/} and that
it is not the right way to regard an anomalous term which is avoidable 
consistently as the anomaly.  We thus conclude that the BRS invariance is 
{\it not\/} violated in the de Donder-gauge two-dimensional quantum gravity.
\par
Then a question arises: Why has Takahashi\cite{Takahashi} obtained $D=26$
precisely by his reasoning on the violation of the BRS invariance?
To answer this, we reformulate Takahashi's calculation so as to be able to
relate it to the discussion made in Sec.\ 2.
\par
First, we assume that the BRS invariance is not violated.  As discussed 
above, the crucial point arises from external lines.  Moreover, external
lines are encountered only when one considers higher-order perturbative
corrections.  That is, it is crucial in Takahashi's calculation to 
consider the term {\it linear\/} in quantum fields, which is always neglected
in the discussions made in Sec.\ 2.  We therefore eliminate the linear
term $b_\rho$ involved in $\tb_\rho$ by  making use of the field equation
\eqno(3,27).  That is, we consider
\begin{eqnarray}
\tilde\calT_{\mu\nu}&\equiv&\calT_{\mu\nu}
  +(\partial_\mu\tb_\nu+\partial_\nu\tb_\mu) \nonumber \\
  &=&\partial_\mu\phi^M\cdot\partial_\nu\phi_M
     -i[\partial_\mu\bar c_\sigma\cdot\partial_\nu c^\sigma
        +\partial_\mu(\partial_\sigma\bar c_\nu\cdot c^\sigma)
         + (\mu\leftrightarrow\nu)] \nonumber\\
   && +\tg_{\mu\nu}\tg^{\sigma\tau}\partial_\sigma b_\tau
      -\tg_{\mu\nu}\tg^{\sigma\tau}\Big({1\over2}\partial_\sigma\phi^M\cdot
    \partial_\tau\phi_M
     -i\partial_\sigma\bar c_\lambda\cdot\partial_\tau c^\lambda\Big)
\end{eqnarray}
instead of $\tb_\nu$.  We then substitute \eqno(2,11) into $\tg^{\mu\nu}$
and neglect higher-order terms:
\begin{eqnarray}
\tilde\calT_{\mu\nu}
&=&\partial_\mu\phi^M\cdot\partial_\nu\phi_M
  -i[\partial_\mu\bar c_\sigma\cdot\partial_\nu c^\sigma
     +\partial_\mu(\partial_\sigma\bar c_\nu\cdot c^\sigma)
   + (\mu\leftrightarrow\nu)]  \nonumber\\
&&+h_{\mu\nu}\partial^\sigma b_\sigma  +\eta_{\mu\nu}\hbox{-terms}.
\end{eqnarray}
This expression is precisely the same as $T^{\hbox{\sc KRBB}}{}_{\mu\nu}$
given in \eqno(2,40)!    Therefore, as we already know, the nonlocal term
of $\wightman{\tilde\calT_{\mu\nu}(x)\tilde\calT_{\lambda\rho}(y)}$ is
proportional to $D-26$.  Thus Takahashi's calculation can be {\it 
reinterpreted\/} as the conformal anomaly of 
$\wightman{\tilde\calT_{\mu\nu}\tilde\calT_{\lambda\rho}}$ in the 
{\it BRS-invariant\/} framework.
\par
Since $\tilde\calT_{\mu\nu}$ consists of $\calT_{\mu\nu}=0$ and a BRS-exact
quantity $\partial_\mu\tb_\nu+\partial_\nu\tb_\mu$, one may wonder why it 
can have the anomaly in the BRS-invariant framework.
The reason is that the field equation $\calT_{\mu\nu}=0$ suffers from the 
field-equation anomaly,\foot(e,{Therefore, those who want to believe neither
BRS violation nor field-equation anomaly cannot explain the conformal
anomaly of $\wightman{\tilde\calT_{\mu\nu}\tilde\calT_{\lambda\rho}}$.
\vskip8pt})
as explained at the end of Sec.\ 3.  
Explicitly, we have\foot(f,{Unfortunately, in Ref.\ 22, $D+8$ is given
instead of $D+10$ in {\eqno(5,7)} because the contribution from 
$g_{\mu\nu}g^{\sigma\tau}\partial_\sigma b_\tau$ has been overlooked there.})
\begin{eqnarray}
&&\wightman{\calT_{\mu\nu}\,\calT_{\lambda\rho}}
 =(D+10)\vPhi_{\mu\nu\lambda\rho}+\cdots, \\
&&\wightman{\calT_{\mu\nu}(\partial_\lambda\tb_\rho+\partial_\rho\tb_\lambda)}
 =-18\vPhi_{\mu\nu\lambda\rho}+\cdots, \\
&&\wightman{(\partial_\mu\tb_\nu+\partial_\nu\tb_\mu)
  (\partial_\lambda\tb_\rho+\partial_\rho\tb_\lambda)}=0,
\end{eqnarray}
so that
\begin{eqnarray}
&&\wightman{\tilde\calT_{\mu\nu}\,\tilde\calT_{\lambda\rho}}
 =[D+10+2\times(-18)]\vPhi_{\mu\nu\lambda\rho}+\cdots.
\end{eqnarray}
Thus we see that the field-equation
anomaly underlies behind the conformal anomaly.
\par
It should be noted further that by setting $D=26$, 
$\wightman{\tilde\calT_{\mu\nu}\tilde\calT_{\lambda\rho}}$ becomes 
anomaly-free, but $\wightman{\tilde\calT_{\mu\nu}{\cal O}}$ generally
remain anomalous.  If, for example, the FP-ghost number current 
$j_{\hbox{\sc c}}{}^\lambda$ is chosen as ${\cal O}$, the coefficient of 
the anomaly term of $\wightman{\tilde\calT_{\mu\nu}j_{\hbox{\sc 
c}}{}^\lambda}$ is known to be $-2$ (independently of 
$D$).\cite{Kraemmer-Rebhan}
\par
On the other hand, as is well known in the Kugo-Ojima 
formalism,\cite{Kugo-Ojima} not only the BRS invariance but also the 
FP-ghost number conservation should not be violated so as to maintain
the physical unitarity of the theory.  As noted in Sec.\ 3, the FP-ghost
number $Q_c$ is {\it not\/} broken in the de Donder-gauge two-dimensional
quantum gravity.  {\it What is really anomalous is the field equation\/}
$\calT_{\mu\nu}=0$.  
\par
To see this fact more clearly, we demonstrate the following proposition.
{\it By using the field-equation anomaly we can eliminate the conformal
anomaly of $\wightman{T_{\mu\nu}\,T_{\lambda\rho}}$ considered in Sec.\ 2\/}, 
provided that no perturbative approximation is made at the Lagrangian
level.  
The decomposition \eqno(2,33) is characterized by the requirements that
$\hat g_{\mu\nu}$ is a $c$-number and that $h_{\mu\nu}$ is independent
of $\hat g_{\mu\nu}$.  Then, as pointed out in Sec.\ 2, 
$\delta/\delta \hat g_{\mu\nu}$ does {\it not\/} commute with the BRS
transformation $\gdel$.  But we can prove that
\begin{eqnarray}
&&\left[ {\delta\over\delta\hat g_{\mu\nu}}
         -{\delta\over\delta g_{\mu\nu}}, \; \gdel \right]=0.
\end{eqnarray}
We {\it redefine\/} ``energy-momentum tensor'' by
\begin{eqnarray}
&&T_{\mu\nu}{}'=-2(-\hat g)^{-1/2}\hat g_{\mu\sigma}\hat g_{\nu\tau}
\left({\delta\over\delta\hat g_{\sigma\tau}}
         -{\delta\over\delta g_{\sigma\tau}}\right)\int d^2x\,\lag
\end{eqnarray}
at the flat limit.  The extra part is proportional to the field equation.
Then $T_{\hbox{\sc GF}\,\mu\nu}{}'+T_{\hbox{\sc FP}\,\mu\nu}{}'$ is 
BRS-exact because of \eqno(5,11) and the BRS exactness of 
$\lagGF+\lagFP$.
Furthermore, the main terms of $T_{\hbox{\sc S}\,\mu\nu}{}'$ are absent
owing to cancellation.  Hence apart from higher-order terms,
$T_{\mu\nu}{}'$ is BRS-exact.  
Thus $\wightman{T_{\mu\nu}{}'T_{\lambda\rho}{}'}$ has no conformal anomaly.
More generally, for any BRS-invariant operator $\calO$, 
$\wightman{T_{\mu\nu}{}'\,\calO}$ has no anomaly.
\par
We thus find that various anomaly-like pathologies of 
the two-dimensional quantum gravity are originated from the 
field-equation anomaly, whose trouble we can bypass {\it without\/} 
adjusting the value of $D$.
\vskip50pt
%
%%%%%%%%%%%%%%%%%%%%%%%%%%%  Section 6  %%%%%%%%%%%%%%%%%%%%%%%%%%%%%
\Sec{Discussion}
The two-dimensional quantum gravity has two faces: The one is the bosonic
string theory, while the other is the $n=2$ version of quantum Einstein
gravity.  The former implies the existence of the critical dimension
$D=26$, while the latter should not have such peculiar feature.
Thus the two-dimensional quantum gravity is something like
``Dr.\ Jekyll and Mr.\ Hyde''.  On the other hand, it is a mathematically
well-defined object at least at the physicist's level of rigor.   
In the present paper, we have investigated how those paradoxical aspects
are reconciled explicitly.
\par
The de Donder gauge two-dimensional quantum gravity is exactly solvable
in the Heisenberg picture.  It is not only the $n=2$ case of the de 
Donder-gauge quantum Einstein gravity but also its zeroth-order 
approximation in the $\kappa$ expansion apart from its dimensionality.
The two-dimensional quantum gravity is a beautiful theory having many
symmetry generators forming $IO\!S\!p(4+D,4)$, none of which is anomalous.
In particular, the BRS invariance and the FP-ghost number conservation 
are unbroken.
\par
However --- a big ``however'' ---, the two-dimensional quantum gravity has the
field-equation anomaly, that is, the field equation \eqno(3,27), i.e.,
\begin{eqnarray}
&&\calT_{\mu\nu}=0,
\end{eqnarray}
is violated at the level of the representation in terms of state vectors.
Taking covariant derivative of \eqno(6,1) and using some other field
equations, one obtains
\begin{eqnarray}
&&\partial_\mu(\tg^{\mu\nu}\partial_\nu b_\rho)=0.
\end{eqnarray}
Although both \eqno(6,1) and \eqno(6,2) have two degrees of freedom,
it is impossible to reproduce \eqno(6,1) from \eqno(6,2) and other
field equations. The healthy results stated above are the consequences
of the framework obtained by replacing \eqno(6,1) by \eqno(6,2).
Thus all the anomalous features of the two-dimensional quantum gravity
are confined into \eqno(6,1) {\it modulo\/} \eqno(6,2).
This part is totally foreign to quantum Einstein gravity.
\par
Now, we proceed to seeing how the characteristic feature of the string
theory arise from the two-dimensional quantum gravity.
As we have seen in Sec.\ 5, the conformal anomaly, the FP-ghost number
current anomaly, etc.\ are the consequence of the field-equation anomaly
for \eqno(6,1).   The key equation is 
\begin{eqnarray}
\tilde\calT_{\mu\nu}&\equiv&\calT_{\mu\nu}
  -i\,\gdel(\partial_\mu\bar c_\nu+\partial_\nu\bar c_\mu) \nonumber \\
  &=&T^{\hbox{\sc KRBB}}{}_{\mu\nu}
\end{eqnarray}
apart from $\eta_{\mu\nu}$-terms and higher-order ones.  Here 
$T^{\hbox{\sc KRBB}}{}_{\mu\nu}$, given in \eqno(2,40), is one of 
``energy-momentum tensors'' derivable by the Kraemmer-Rebhan procedure.
The two-point function of it gives the conformal anomaly proportional
to $D-26$ according to the gauge independence of the conformal anomaly
in the sense of Kraemmer and Rebhan.
\par
The critical dimension $D=26$ is obtained as long as one follows the 
Kraemmer-Rebhan procedure, which is characterized by adopting $\tb_\rho$
as the primary field and by employing perturbative approach from the 
outset.  As shown in Sec.\ 4, the two-point function of ``energy-momentum
tensor'' $T_{\mu\nu}$ no longer gives $D=26$ if the above setting-up is 
abandoned.  Thus the critical dimension $D=26$ is {\it not\/} the 
indispensable consequence of the two-dimensional quantum gravity, but
a consequence of a {\it particular approach to it\/}.
That is, the string theory is the two-dimensional quantum gravity 
{\it plus something\/} (choice of a particular perturbation theory).
This consideration suggests that it is impossible to formulate the 
string theory covariantly in the Heisenberg picture.
\vfill\eject
%
%%%%%%%%%%%%%%%%%%%%%%%%%%%%   References  %%%%%%%%%%%%%%%%%%%%%%%%%%%%%

\end{document}